\documentclass[twocolumn,fp,a4paper]{jpsj3}

\usepackage{subfigure}
\usepackage{wrapfig}
\usepackage{color}
\usepackage{graphicx}
\usepackage{dcolumn}
\usepackage{bm}
\usepackage{txfonts}

\title{Axionic Antiferromagnetic Insulator Phase in a Correlated and Spin-Orbit Coupled System}

\author{\name{Akihiko \surname{Sekine}}\thanks{E-mail: sekine@imr.tohoku.ac.jp} and \name{Kentaro \surname{Nomura}}}
\inst{Institute for Materials Research, Tohoku University, Sendai 980-8577, Japan} 

\abst{We study theoretically a three-dimensional correlated and spin-orbit coupled system, the half-filled extended Fu-Kane-Mele-Hubbard model on a diamond lattice, focusing on the topological magnetoelectric response of the antiferromagnetic insulator phase.
In the antiferromagnetic insulator phase, the Dirac-like low-energy effective Hamiltonian is obtained.
Then the theta term, which results in the magnetoelectric response, is derived as a consequence of the chiral anomaly.
The realization of the dynamical axion field in our model is discussed.
The relation with a symmetry broken phase induced by interactions in lattice quantum chromodynamics is also discussed.}

\begin{document}
\maketitle

\section{Introduction}
A great number of studies on topological insulators have been done since the pioneering works \cite{Kane2005} appeared, in the search for novel phenomena due to the topological properties of the system.
The most prominent feature common to two-dimensional (2D) and 3D topological insulators is the existence of the edge (surface) states which are protected by time-reversal symmetry.
These edge or surface states are known to be robust against perturbations.
On the one hand, one of the noteworthy characters peculiar to 3D topological insulators is the topological magnetoelectric effect which is described by the theta term \cite{Qi2008}.
The theta term is written as
\begin{align}
\begin{split}
S_\theta=\int dtd^3 x \frac{\theta e^2}{4\pi^2\hbar c}\bm{E}\cdot\bm{B},\label{S_theta_realtime}
\end{split}
\end{align}
where $\bm{E}$ and $\bm{B}$ are an electric field and magnetic field, respectively.
From this action, we obtain the cross-correlated responses expressed by $\bm{P}=\theta e^2/(4\pi^2\hbar c)\bm{B}$ and $\bm{M}=\theta e^2/(4\pi^2\hbar c)\bm{E}$, with $\bm{P}$ the electric polarization and $\bm{M}$ the magnetization.

In the field theory literature, the action (\ref{S_theta_realtime}) is termed the axion electrodynamics \cite{Wilczek1987}.
The axion is an elementary particle proposed about forty years ago to solve the so-called strong CP problem in quantum chromodynamics (QCD) \cite{Peccei1977,Weinberg1978,Wilczek1978}.
By subsequent studies, the axion is now considered to be essential to explain experimental results in particle physics and astrophysics \cite{book-axions}.
The axion is also considered as a candidate for dark matter \cite{book-axions}.
However, regardless of intensive experimental searches, the axion has not yet been found.
The axions interact with photons, and the axion-photon coupling is described by Eq. (\ref{S_theta_realtime}) with $\theta$ being the axion field.
Therefore, observing the magnetoelectric responses originating from Eq. (\ref{S_theta_realtime}) in condensed matter is equivalent to realizing the (dynamical) axion field.
There have been some theoretical studies which propose ways to observe experimentally the dynamical axion field in condensed matter \cite{Li2010,Ooguri2012}.

When the system is time-reversal invariant, the condition that $\theta=\pi$ (mod $2\pi$) is imposed for 3D topological insulators, and $\theta=0$ for normal insulators.
On the other hand, when time-reversal symmetry of the system is broken, the value of $\theta$ can be arbitrary.
In general, the value of $\theta$ can be calculated according to the formula \cite{Qi2008}
\begin{align}
\begin{split}
\theta=\frac{1}{4\pi}\int_{\rm BZ}d^3k\epsilon^{ijk}{\rm Tr}\left[\mathcal{A}_i\partial_j\mathcal{A}_k-i\frac{2}{3}\mathcal{A}_i\mathcal{A}_j\mathcal{A}_k\right],\label{exact-theta1}
\end{split}
\end{align}
where $\mathcal{A}^{\mu\nu}_j=i\langle u_\mu|\partial/\partial k_j|u_\nu\rangle$, and $|u_\nu\rangle$ is the periodic Bloch function with $\nu$ the occupied bands.
We can calculate $\theta$ from other equivalent expressions \cite{Essin2009,Essin2010,Coh2011}.
However, some techniques (such as choosing a gauge for $\mathcal{A}$) are required to calculate numerically.
In systems where the single-particle Hamiltonian can be described as $\mathcal{H}(\bm{k})=\sum_{\mu=1}^5R_\mu(\bm{k})\alpha_\mu$ with matrices $\alpha_\mu$ satisfying the Clifford algebra $\{\alpha_\mu,\alpha_\nu\}=2\delta_{\mu\nu}$, there exists an explicit expression for $\theta$
\cite{Li2010,Wang2011}:
\begin{align}
\begin{split}
\theta&=\frac{1}{4\pi}\int_{\rm BZ} d^3 k \frac{2|R|+R_4}{(|R|+R_4)^2|R|^3}\epsilon^{ijkl}R_i\frac{\partial R_j}{\partial k_x}\frac{\partial R_k}{\partial k_y}\frac{\partial R_l}{\partial k_z},\label{exact-theta2}
\end{split}
\end{align}
where $i,j,k,l=1,2,3,5$ and $|R|=\sqrt{\sum_{\mu=1}^5R_\mu^2}$.
Here note that only the matrix $\alpha_4$ is even under time-reversal.
In this work, we derive an analytical expression for $\theta$ in a time-reversal symmetry broken phase with the use of a field-theoretical method.

The spin-orbit interaction has been revealed to be important to realize topologically nontrivial phases.
On the other hand, the effects of the electron-electron interaction has been a central subject in modern condensed matter physics.
Now, the interplay of spin-orbit coupling and electron correlation is a hot topic.
One of the triggers is the discovery of a novel Mott insulating state in a correlated $5d$ electron system, which revealed that the Mott insulating state is induced by strong spin-orbit coupling \cite{Kim2008,Kim2009}.
Recent studies have shown that novel phases, such as the quantum spin Hall insulator \cite{Shitade2009}, the topological Mott insulator \cite{Pesin2010}, the topological magnetic insulator \cite{Li2010,Wang2011}, the Weyl semimetal \cite{Wan2011}, and the phase which is a condensed-matter analog of a phase in lattice QCD \cite{Sekine2013}, emerge by the interplay of spin-orbit coupling and electron correlation.

Electron correlation effects in topological insulators have also been investigated intensively \cite{Hohenadler2013}.
The Kane-Mele model on the honeycomb lattice is well known as a model which describes a 2D topological insulator \cite{Kane2005}.
The Kane-Mele model with on-site interaction, the Kane-Mele-Hubbard model, has been one of the most investigated system so far.
In this system, the antiferromagnetic insulator phase develops in the region where the on-site interaction strength $U$ is strong \cite{Rachel2010,Hohenadler2011,Yamaji2011,Hohenadler2012,Vaezi2012,Wu2012}.
When the strength $U$ is intermediate, it has been shown that the spin liquid phase emerges \cite{Hohenadler2011,Yamaji2011,Hohenadler2012,Vaezi2012,Wu2012} and pointed out the possibility of the fractional topological insulator phase \cite{Rachel2010}.
In another model of a 2D topological insulator with on-site interaction, the Bernevig-Hughes-Zhang-Hubbard model, the existence of the topological antiferromagnetic insulator phase has been pointed out \cite{Yoshida2013}.
On the other hand, in the case of three-dimensions, the Fu-Kane-Mele model on the diamond lattice, the 3D analog of the Kane-Mele model, is known as a model for a 3D topological insulator \cite{Fu2007,Fu2007a}.
What is the properties of an interacting Fu-Kane-Mele model, the Fu-Kane-Mele-Hubbard model?
So far there has been no study on this model, although interesting phenomena are expected to emerge.

In this paper, we focus on the topological magnetoelectric response of the antiferromagnetic insulator phase in the extended Fu-Kane-Mele-Hubbard model on a diamond lattice at half-filling, within the mean-field approximation.
This paper is organized as follows.
In Sec. \ref{model}, the model we adopt is explained.
We take into account the on-site and nearest-neighbor repulsive electron-electron interactions.
In Sec. \ref{phasediagram}, the mean-field phase diagram is presented.
In Sec. \ref{me-response}, we obtain analytically the value of $\theta$ in the antiferromagnetic insulator phase.
First we show that we can derive the Dirac Hamiltonian in the antiferromagnetic insulator phase.
Then based on the Fujikawa's method \cite{Fujikawa1979,Fujikawa1980}, we obtain the theta term as a consequence of the chiral anomaly.
In Sec. \ref{discussions}, we discuss the realization of the dynamical axion field in our model.
we also discuss the relation between our antiferromagnetic insulator phase and the so-called ``Aoki phase'', a symmetry broken phase induced by interactions in lattice QCD \cite{Aoki1984}.

\section{Model \label{model}}
\begin{figure}[!t]
\centering
\includegraphics[width=1.0\columnwidth,clip]{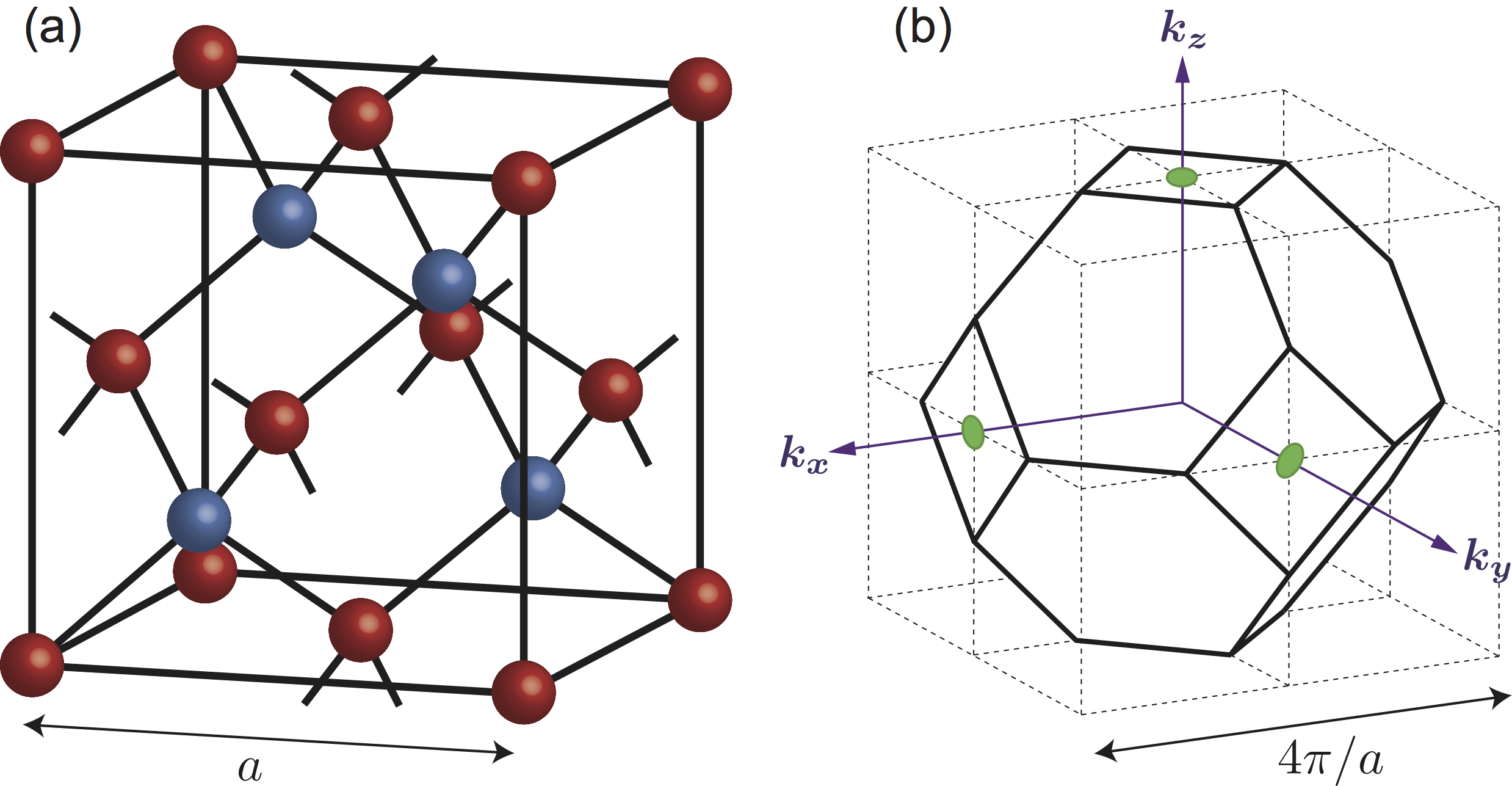}
\caption{(Color online) (a) A diamond lattice, which consists of two sublattices (red and blue), and each sublattice forms a fcc lattice. (b) The first Brillouin zone of a fcc lattice. Green circles represent the $X$ points.}\label{Fig1}
\end{figure}

Let us consider a 3D lattice model with electron correlation and spin-orbit coupling.
The model we adopt is the extended Fu-Kane-Mele-Hubbard model on a diamond lattice at half-filling, in which the Hamiltonian is given by $H=H_0+H_{\rm int}$ with the non-interacting part
\begin{align}
\begin{split}
H_0&=\sum_{\langle i,j\rangle,\sigma}t_{ij}c^\dag_{i\sigma}c_{j\sigma}+i\frac{4\lambda}{a^2}\sum_{\langle\langle i,j\rangle\rangle}c^\dag_{i}\bm{\sigma}\cdot(\bm{d}^1_{ij}\times\bm{d}^2_{ij})c_{j},\end{split}
\end{align}
and the interaction part
\begin{align}
\begin{split}
H_{\rm int}&=U\sum_i n_{i\uparrow}n_{i\downarrow}+\sum_{\langle i,j\rangle}V_{ij}n_i n_j,\label{hamiltonian}
\end{split}
\end{align}
where $c^\dag_{i\sigma}$ is an electron creation operator at a site $i$ with spin $\sigma(=\uparrow,\downarrow)$, $n_{i\sigma}=c^\dag_{i\sigma}c_{i\sigma}$, $n_i=n_{i\uparrow}+n_{i\downarrow}$, and $a$ is the lattice constant of the fcc lattice.
The first and second terms of $H_0$ represent the nearest-neighbor hopping and the next-nearest-neighbor spin-orbit coupling, respectively.
$\bm{d}^1_{ij}$ and $\bm{d}^2_{ij}$ are the two vectors which connect two sites $i$ and $j$ of the same sublattice.
They are given by two of the four nearest-neighbor vectors, $\frac{a}{4}(1,1,1)$, $\frac{a}{4}(-1,-1,1)$, $\frac{a}{4}(1,-1,-1)$, and $\frac{a}{4}(-1,1,-1)$,  with proper signs (directions of the vectors).
$\bm{\sigma}=(\sigma_1,\sigma_2,\sigma_3)$ are the Pauli matrices for the spin degree of freedom.
The first and second terms of $H_{\rm int}$ describe the on-site and nearest-neighbor repulsive electron-electron interactions, respectively.
The lattice structure of a diamond lattice is shown in Fig. \ref{Fig1}(a).

It is convenient to express the non-interacting part $H_0$ of the Hamiltonian in terms of the 4$\times$4 alpha (gamma) matrices. 
The diamond lattice consists of two sublattices ($A$ and $B$), with each sublattice forming a fcc lattice.
In such a case, we can define the basis $c_{\bm{k}}\equiv[c_{\bm{k}A\uparrow},c_{\bm{k}A\downarrow},c_{\bm{k}B\uparrow},c_{\bm{k}B\downarrow}]^T$   where the wave vector $\bm{k}$ is given by the points in the first Brillouin zone of the fcc lattice [see Fig. \ref{Fig1}(b)].
Then the single-particle Hamiltonian $\mathcal{H}_0(\bm{k})$ [$H_0\equiv\sum_{\bm{k}}c^\dag_{\bm{k}}\mathcal{H}_0(\bm{k})c_{\bm{k}}$] is written as \cite{Fu2007,Fu2007a}
\begin{align}
\begin{split}
\mathcal{H}_0(\bm{k})=\sum_{\mu=1}^5R_\mu(\bm{k})\alpha_\mu,\label{H_0-singleparticle}
\end{split}
\end{align}
where the coefficients $R_\mu(\bm{k})$ are given by
\begin{align}
\begin{split}
R_1(\bm{k})&=\lambda[\sin u_2-\sin u_3-\sin(u_2-u_1)+\sin(u_3-u_1)],\\
R_2(\bm{k})&=\lambda[\sin u_3-\sin u_1-\sin(u_3-u_2)+\sin(u_1-u_2)],\\
R_3(\bm{k})&=\lambda[\sin u_1-\sin u_2-\sin(u_1-u_3)+\sin(u_2-u_3)],\\
R_4(\bm{k})&=t+\delta t_1+t(\cos u_1+\cos u_2+\cos u_3),\\
R_5(\bm{k})&=t(\sin u_1+\sin u_2+\sin u_3).\\
\end{split}
\end{align}
Here $u_1=\bm{k}\cdot\bm{a}_1$, $u_2=\bm{k}\cdot\bm{a}_2$, and $u_3=\bm{k}\cdot\bm{a}_3$ with
$\bm{a}_1=\frac{a}{2}(0,1,1)$, $\bm{a}_2=\frac{a}{2}(1,0,1)$ and $\bm{a}_3=\frac{a}{2}(1,1,0)$ being the primitive translation vectors.
In the following, we set $a=1$.
The alpha matrices $\alpha_\mu$ are given by the chiral representation: 
\begin{align}
\begin{split}
\alpha_j=
\begin{bmatrix}
\sigma_j & 0\\
0 & -\sigma_j
\end{bmatrix},\ \ \ 
\alpha_4=
\begin{bmatrix}
0 & 1\\
1 & 0
\end{bmatrix},\ \ \ 
\alpha_5=
\begin{bmatrix}
0 & -i\\
i & 0
\end{bmatrix},
\end{split}
\end{align}
where $j=1,2,3$.
In the present basis, the time-reversal operator and spatial inversion (parity) operator are given by $\mathcal{T}=\bm{1}\otimes(-i\sigma_2)\mathcal{K}$ ($\mathcal{K}$ is the complex conjugation operator) and $\mathcal{P}=\tau_1\otimes\bm{1}$, respectively.
We have introduced the hopping strength anisotropy $\delta t_1$ due to the lattice distortion along the [111] direction.
Namely, we have set such that $t_{ij}=t+\delta t_1$ for the [111] direction, and $t_{ij}=t$ for the other three directions.
When $\delta t_1=0$, the system is a semimetal, i.e., the energy bands touch at the three points $X^r=2\pi (\delta_{rx},\delta_{ry},\delta_{rz})$ ($r=x,y,z$).
Finite $\delta t_1$ opens a gap of $2|\delta t_1|$ at the $X^r$ points.

The $Z_2$ invariant of the system is given by
\begin{align}
\begin{split}
(-1)^{\nu_0}=\prod_{i=1}^8{\rm sgn}\left[t+\delta t_1+t\sum_{p=1}^3\cos\left(\bm{\Gamma}_i\cdot\bm{a}_p\right)\right],
\end{split}
\end{align}
where $\bm{\Gamma}_i$ are the eight time-reversal invariant momenta: $(0,0,0)$, $(2\pi,0,0)$, $(0,2\pi,0)$, $(0,0,2\pi)$, $(\pi,\pi,\pi)$, $(\pi,\pi,-\pi)$, $(\pi,-\pi,\pi)$, and $(-\pi,\pi,\pi)$.
We see that the system is a topological insulator (normal insulator) when $0<\delta t_1<2t$ ($\delta t_1<0$ or $\delta t_1>2t$).
Note that in this paper we do not distinguish a weak topological insulator from a normal insulator.

Let us look at closely $\mathcal{H}_0(\bm{k})$ around the $X^r$ points.
Setting $\bm{k}=X^r+\bm{q}$ and retaining the terms up to the order of $q$, we obtain the low-energy effective Hamiltonian near the Fermi level around each $X$ point \cite{Fu2007,Fu2007a}:
\begin{align}
\begin{split}
\mathcal{H}_0(X^x+\bm{q})&=tq_x\alpha_5+2\lambda q_y\alpha_2-2\lambda q_z\alpha_3+\delta t_1\alpha_4,\\
\mathcal{H}_0(X^y+\bm{q})&=tq_y\alpha_5+2\lambda q_z\alpha_3-2\lambda q_x\alpha_1+\delta t_1\alpha_4,\\
\mathcal{H}_0(X^z+\bm{q})&=tq_z\alpha_5+2\lambda q_x\alpha_1-2\lambda q_y\alpha_2+\delta t_1\alpha_4.\label{low-energy-H0}
\end{split}
\end{align}
These are so-called the Dirac Hamiltonian.
For example, the energy spectrum around the $X^x$ point is readily obtained as
\begin{align}
\begin{split}
E(X^x+\bm{q})=\pm\sqrt{(tq_x)^2+(2\lambda q_y)^2+(2\lambda q_z)^2+(\delta t_1)^2}.
\end{split}
\end{align}
As mentioned above, we see that the system is gapless when $\delta t_1=0$ and nonzero $\delta t_1$ is regarded as the mass of the Dirac quasiparticles.
At each $X^r$ point, one of the three components which originate from spin-orbit coupling $R_r(\bm{k})$ disappears and instead $R_5(\bm{k})$ compensates for the $q_r$-dependence of the effective Hamiltonian.

\section{Mean-field phase diagram \label{phasediagram}}
{\it Spin-density wave instability.---} Let us perform the mean-field approximation to the interaction term and derive the mean-field Hamiltonian of the system.
First we consider the spin-density wave (SDW) instability.
To do this, we firstly approximate the on-site interaction $H_U=U\sum_i n_{i\uparrow}n_{i\downarrow}$ as
\begin{align}
\begin{split}
H_U&\approx  U\sum_i \left[ \langle n_{i\downarrow}\rangle n_{i\uparrow}+\langle n_{i\uparrow}\rangle n_{i\downarrow}-\langle n_{i\uparrow}\rangle\langle n_{i\downarrow}\rangle\right.\\
&\quad\left.-\langle c^\dagger_{i\uparrow}c_{i\downarrow}\rangle c^\dagger_{i\downarrow}c_{i\uparrow}-\langle c^\dagger_{i\downarrow}c_{i\uparrow}\rangle c^\dagger_{i\uparrow}c_{i\downarrow}+\langle c^\dagger_{i\uparrow}c_{i\downarrow}\rangle \langle c^\dagger_{i\downarrow}c_{i\uparrow}\rangle \right].
\end{split}
\end{align}
Due to the spin-orbit coupling, the spin SU(2) symmetry is broken and the orientations of the spins are coupled to the lattice structure.
We assume the antiferromagnetic ordering between the two sublattices in terms of the spherical coordinate $(m,\theta,\varphi)$:
\begin{align}
\begin{split}
\langle \bm{S}_{i'A}\rangle=-\langle \bm{S}_{i'B}\rangle&=(m\sin\theta\cos\varphi, m\sin\theta\sin\varphi, m\cos\theta)\\
&\equiv m_1\bm{e}_x+m_2\bm{e}_y+m_3\bm{e}_z,\label{AF-order}
\end{split}
\end{align}
where $\langle \bm{S}_{i'\mu}\rangle=\frac{1}{2}\langle c^\dagger_{i'\mu\alpha}\bm{\sigma}_{\alpha\beta}c_{i'\mu\beta}\rangle$ $(\mu=A,B)$ with $i'$ denoting the $i'$-th unit cell.
Then after a calculation, we obtain
\begin{align}
\begin{split}
U\sum_i \left[ -\langle n_{i\uparrow}\rangle\langle n_{i\downarrow}\rangle +\langle c^\dagger_{i\uparrow}c_{i\downarrow}\rangle \langle c^\dagger_{i\downarrow}c_{i\uparrow}\rangle \right]=2NU \sum_f m^2_f,\label{MF-1}
\end{split}
\end{align}
\begin{align}
\begin{split}
&U\sum_i \left[ \langle n_{i\downarrow}\rangle n_{i\uparrow}+\langle n_{i\uparrow}\rangle n_{i\downarrow}-\langle c^\dagger_{i\uparrow}c_{i\downarrow}\rangle c^\dagger_{i\downarrow}c_{i\uparrow}-\langle c^\dagger_{i\downarrow}c_{i\uparrow}\rangle c^\dagger_{i\uparrow}c_{i\downarrow} \right]\\
&=-U\sum_{i'}\sum_{f=1,2,3}c^\dagger_{i'A}[m_f\sigma_f] c_{i'A}+U\sum_{i'}\sum_{f=1,2,3}c^\dagger_{i'B}[m_f\sigma_f]c_{i'B}\\
&=-U\sum_{\bm{k}}c^\dagger_{\bm{k}}[m_1\alpha_1+m_2\alpha_2+m_3\alpha_3]c_{\bm{k}},\label{MF-2}
\end{split}
\end{align}
where $N$ is the number of the unit cells and the wave vectors $\bm{k}$ take $N$ points in the first Brillouin zone of the fcc lattice.
This equation means that the on-site interaction term has the same matrix form as the spin-orbit interaction term in the mean-field level.
A similar result has been obtained in the Kane-Mele-Hubbard model on a honeycomb lattice \cite{Rachel2010}.
Here we have omitted irrelevant constant terms in Eqs. (\ref{MF-1}) and (\ref{MF-2}).

Secondly, we approximate the nearest-neighbor interaction $H_V=\sum_{\langle i,j\rangle}V_{ij}n_i n_j$ as
\begin{align}
\begin{split}
H_V&\approx  -\sum_{\langle i,j\rangle}\sum_{\sigma,\sigma'} V_{ij}\left[\langle c^\dagger_{i\sigma}c_{j\sigma'}\rangle c^\dagger_{j\sigma'}c_{i\sigma}+\langle c^\dagger_{j\sigma'}c_{i\sigma}\rangle c^\dagger_{i\sigma}c_{j\sigma'}\right.\\
&\quad\left.-\langle c^\dagger_{i\sigma}c_{j\sigma'}\rangle \langle c^\dagger_{j\sigma'}c_{i\sigma}\rangle \right].
\end{split}
\end{align}
We assume that the values of $\langle c^\dagger_{i\sigma}c_{j\sigma'}\rangle$ depend on the hopping strength, namely we set $\langle c^\dagger_{i\sigma}c_{j\sigma'}\rangle=-\Delta\delta_{\sigma\sigma'} t_{ij}/t$.
On the other hand, we neglect the interaction strength anisotropy due to the lattice distortion for simplicity, i.e., we set $V_{ij}=V$.
This does not change the resulting phase diagram qualitatively.
After a calculation, we obtain
\begin{align}
\begin{split}
H_V^{\rm MF}=2NV\left[3+(1+\delta t_1/t)^2\right]\Delta^2+V\Delta/t\sum_{\langle i,j\rangle,\sigma}t_{ij} c^\dagger_{i\sigma}c_{j\sigma}.\label{MF-3}
\end{split}
\end{align}

Finally combining Eqs. (\ref{H_0-singleparticle}), (\ref{MF-1}), (\ref{MF-2}), and (\ref{MF-3}), the mean-field Hamiltonian of the system is given by
\begin{align}
\begin{split}
H^{\rm MF}_{\rm SDW}&=2NUm^2+2NV\left[3+(1+\delta t_1/t)^2\right]\Delta^2\\
&\quad+\sum_{\bm{k}}c^\dagger_{\bm{k}}\left[\sum_{\mu=1}^5\tilde{R}_\mu(\bm{k})\alpha_\mu\right]c_{\bm{k}},\label{SDW-Hamiltonian}
\end{split}
\end{align}
where $\tilde{R}_1(\bm{k})=R_1(\bm{k})-Um_1$, $\tilde{R}_2(\bm{k})=R_2(\bm{k})-Um_2$, $\tilde{R}_3(\bm{k})=R_3(\bm{k})-Um_3$, $\tilde{R}_4(\bm{k})=(1+V\Delta/t)R_4(\bm{k})$, and $\tilde{R}_5(\bm{k})=(1+V\Delta/t)R_5(\bm{k})$.
Note that $m_1^2+m_2^2+m_3^2=m^2$.
The free energy at zero temperature for the SDW instability is readily obtained as
\begin{align}
\begin{split}
F_{\rm SDW}(m, \theta, \varphi, \Delta)&=2NUm^2+2NV\left[3+(1+\delta t_1/t)^2\right]\Delta^2\\
&\quad-2\sum_{\bm{k}}\sqrt{{\sum}_{\mu=1}^5\left[\tilde{R}_\mu(\bm{k})\right]^2}.\label{F-SDW}
\end{split}
\end{align}

{\it Charge-density wave instability.---} Next we consider the charge-density wave (CDW) instability.
To do this, we approximate the interaction terms $H_U$ and $H_V$ as
\begin{align}
\begin{split}
H_U&\approx U\sum_i \left[ \langle n_{i\downarrow}\rangle n_{i\uparrow}+\langle n_{i\uparrow}\rangle n_{i\downarrow}-\langle n_{i\uparrow}\rangle\langle n_{i\downarrow}\rangle\right],
\end{split}
\end{align}
\begin{align}
\begin{split}
H_V&\approx \sum_{\langle i,j\rangle}V_{ij} \left\{ \langle n_i\rangle n_j+\langle n_j\rangle n_i-\langle n_i\rangle\langle n_j\rangle-\sum_{\sigma,\sigma'}\left[\langle c^\dagger_{i\sigma}c_{j\sigma'}\rangle c^\dagger_{j\sigma'}c_{i\sigma}\right.\right.\\
&\quad \left.\left.+\langle c^\dagger_{j\sigma'}c_{i\sigma}\rangle c^\dagger_{i\sigma}c_{j\sigma'}
-\langle c^\dagger_{i\sigma}c_{j\sigma'}\rangle \langle c^\dagger_{j\sigma'}c_{i\sigma}\rangle \right]\right\}.
\end{split}
\end{align}
We assume a charge imbalance between the two sublattices such that $\langle n_{i'A\sigma}\rangle=(1+\rho)/2$ and$\langle n_{i'B\sigma}\rangle=(1-\rho)/2$.
As for $H_V$, we assume $\langle c^\dagger_{i\sigma}c_{j\sigma'}\rangle=-\Delta\delta_{\sigma\sigma'} t_{ij}/t$ and $V_{ij}=V$ as in the case of SDW instability.
Then the mean-field Hamiltonian of the interaction term is obtained as
\begin{align}
\begin{split}
H_{\rm int}^{\rm MF}&=NC\rho^2+2NV\left[3+(1+\delta t_1/t)^2\right]\Delta^2\\
&\quad-C\rho\sum_{\bm{k}}c^\dag_{\bm{k}}(\tau_3\otimes\bm{1})c_{\bm{k}}+V\Delta/t\sum_{\langle i,j\rangle,\sigma}t_{ij} c^\dagger_{i\sigma}c_{j\sigma},\label{MF-CDW}
\end{split}
\end{align}
where $C=4V-U/2$, $N$ is the number of the unit cells, and the wave vectors $\bm{k}$ take $N$ points in the first Brillouin zone of the fcc lattice.
Combining Eqs. (\ref{H_0-singleparticle}) and (\ref{MF-CDW}), we obtain the mean-field Hamiltonian of the system.
The matrix $\tau_3\otimes\bm{1}$ is different from the alpha matrices $\alpha_\mu$, and thus the free energy  for the CDW instability is a little complicated but can be obtained analytically as
\begin{align}
\begin{split}
F_{\rm CDW}(\rho, \Delta)&=NC\rho^2+2NV\left[3+(1+\delta t_1/t)^2\right]\Delta^2\\
&\quad-\sum_{\bm{k}}\sum_{\epsilon=\pm1}\sqrt{{\tilde{R}}^2+C^2\rho^2+2C\rho\epsilon \sqrt{\gamma^2}},\label{F-CDW}
\end{split}
\end{align}
where $\tilde{R}^2={\sum}_{\mu=1}^5[\tilde{R}_\mu(\bm{k})]^2$ and $\gamma^2={\sum}_{j=1}^3[\tilde{R}_j(\bm{k})]^2$ with $\tilde{R}_1(\bm{k})=R_1(\bm{k})$, $\tilde{R}_2(\bm{k})=R_2(\bm{k})$, $\tilde{R}_3(\bm{k})=R_3(\bm{k})$, $\tilde{R}_4(\bm{k})=(1+V\Delta/t)R_4(\bm{k})$, and $\tilde{R}_5(\bm{k})=(1+V\Delta/t)R_5(\bm{k})$.

\begin{figure}[!t]
\centering
\includegraphics[width=0.95\columnwidth,clip]{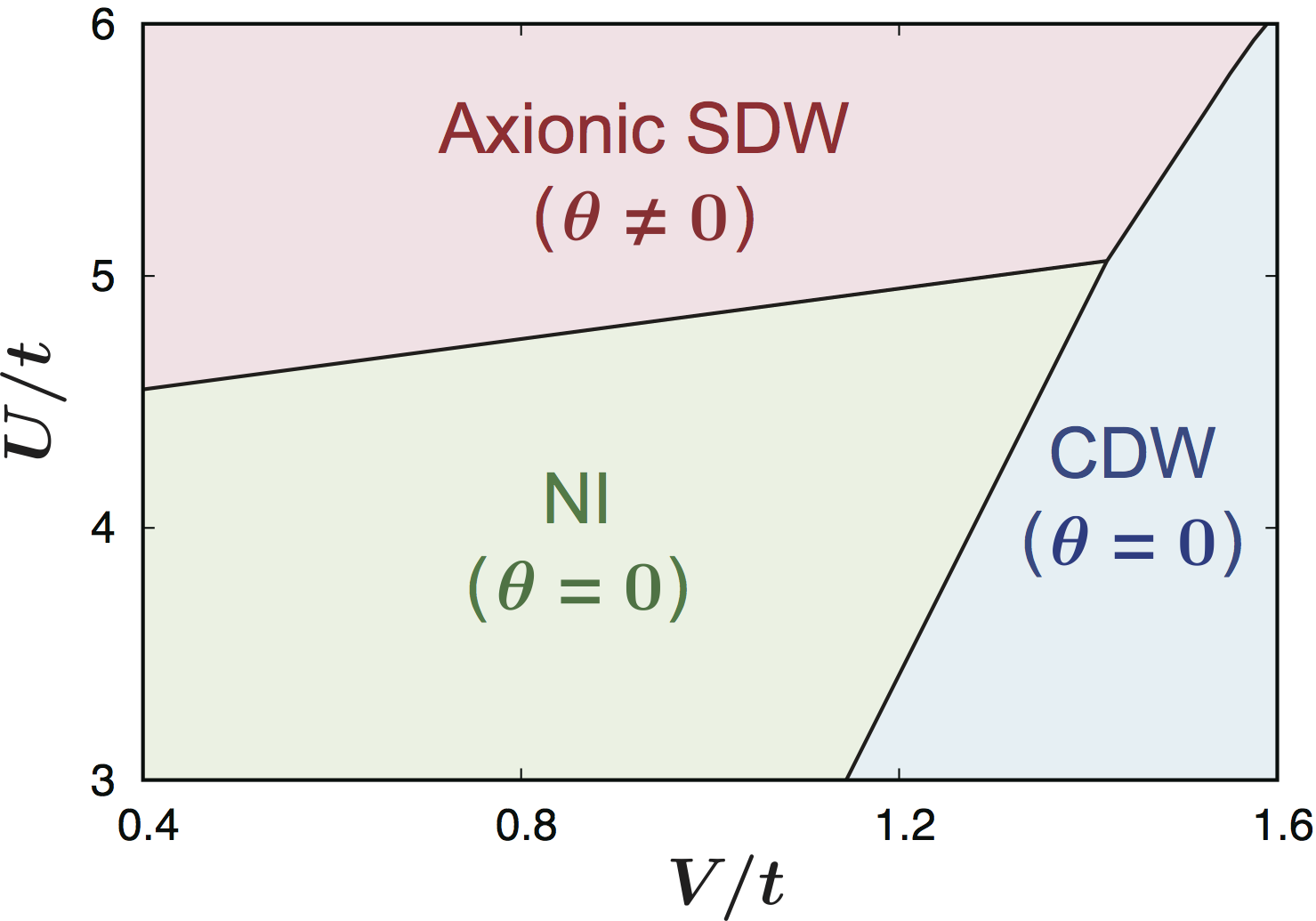}
\caption{(Color online) Mean-field phase diagram of the extended Fu-Kane-Mele-Hubbard model at half-filling.
The strengths of the spin-orbit interaction and lattice distortion are $\lambda/t=0.4$ and $\delta t_1/t=-0.4$, respectively.
The antiferromagnetic ordering is set to be the [111] direction as an example.
The spin-density wave (SDW) and charge-density wave (CDW) phases are given by nonzero $m$ and $\rho$, respectively. When $m=\rho=0$, the system is a normal insulator (NI).
In all the phases, $\Delta$ take nonzero positive values.
In the SDW phase, the topological magnetoelectric response described by the theta term [Eq. (\ref{S_theta_realtime})] arises.}\label{Fig2}
\end{figure}
{\it Mean-field phase diagram.---} To obtain the mean-field ground-state phase diagram, we have to minimize the free energies (\ref{F-SDW}) and (\ref{F-CDW}) by the conditions $\partial F_{\rm SDW}/\partial m=\partial F_{\rm SDW}/\partial \theta=\partial F_{\rm SDW}/\partial \varphi=\partial F_{\rm SDW}/\partial \Delta=0$ and $\partial F_{\rm CDW}/\partial \rho=\partial F_{\rm CDW}/\partial \Delta=0$, and then we have to compare them.
The phase diagram with the antiferromagnetic ordering set to be the [111] direction is shown in Fig. \ref{Fig2} as an example.
The phase diagrams for the other directions and for the positive $\delta t_1$ are qualitatively the same as Fig. \ref{Fig2}.
It was found that the transition from the normal insulator (or topological insulator) phase to the SDW phase is of the second-order, and that the transition from the topological insulator (or normal insulator) phase to the CDW phase is of the first-order.
The values of $\Delta$ are always nonzero and positive when $V\neq 0$.
The obtained phase diagram looks similar to those of conventional correlated electron systems (i.e., the Hubbard models) \cite{Zhang-Kurita}.
Namely, strong on-site electron-electron interaction induces SDW phase and strong nearest-neighbor electron-electron interaction induces CDW phase.
However, note that other phases might be found when our model is studied beyond the mean-field approximation.
Actually, another phase has been reported between the SDW and CDW phases by studies beyond the mean-field approximation, for example, in the half-filled one-dimensional extended Hubbard model \cite{Nakamura,Tsuchiizu2002}.
As is shown later, what is different from usual systems in our model is that the topological magnetoelectric response due to the existence of the theta term can arise in the SDW phase.
In this sense, we call the SDW phase in our model the ``axionic SDW'' (or the ``axionic antiferromagnetic insulator'').
For the purpose of this study, that we derive the theta term in a time-reversal symmetry broken phase, we focus on the SDW phase in the following.

\section{Magnetoelectric response of the antiferromagnetic insulator phase \label{me-response}}
{\it Low-energy effective Hamiltonian.---} Let us investigate the properties of the SDW phase, namely the antiferromagnetic insulator phase.
We consider the general case characterized by the order parameter (\ref{AF-order}).
When $Um_f\ll 2\lambda$ ($f=1,2,3$), we can derive the Dirac Hamiltonian around the $\tilde{X}^r$ points which are slightly deviated from the $X^r$ points:
\begin{align}
\begin{split}
\mathcal{H}(\tilde{X}^x+\bm{q})&=t'q_x\alpha_5+2\lambda q_y\alpha_2-2\lambda q_z\alpha_3+\delta t_1'\alpha_4-Um_1\alpha_1,\\
\mathcal{H}(\tilde{X}^y+\bm{q})&=t'q_y\alpha_5+2\lambda q_z\alpha_3-2\lambda q_x\alpha_1+\delta t_1'\alpha_4-Um_2\alpha_2,\\
\mathcal{H}(\tilde{X}^z+\bm{q})&=t'q_z\alpha_5+2\lambda q_x\alpha_1-2\lambda q_y\alpha_2+\delta t_1'\alpha_4-Um_3\alpha_3,\label{H(k)}
\end{split}
\end{align}
where $t'=t(1+V\Delta/t)$, $\delta t_1'=\delta t_1(1+V\Delta/t)$, $\tilde{X}^x=\left(2\pi, \frac{Um_2}{2\lambda}, -\frac{Um_3}{2\lambda}\right)$, $\tilde{X}^y=\left(-\frac{Um_1}{2\lambda}, 2\pi, \frac{Um_3}{2\lambda}\right)$, and $\tilde{X}^z=\left(\frac{Um_1}{2\lambda}, -\frac{Um_2}{2\lambda}, 2\pi\right)$.
For example, the energy spectrum around the $\tilde{X}^x$ point is readily obtained as
\begin{align}
\begin{split}
&E(\tilde{X}^x+\bm{q})\\&=\pm\sqrt{(t'q_x)^2+(2\lambda q_y)^2+(2\lambda q_z)^2+(\delta t_1')^2+(Um_1)^2}.
\end{split}
\end{align}
We see from Eq. (\ref{H(k)}) that the antiferromagnetic ordering opens a gap at the $\tilde{X}^r$ points, i.e., lowers the energy of the system.
When $Um_f$ is not small compared to $2\lambda$, it is not apparent that the Dirac Hamiltonian can be derived.
Thus in the following, we assume that $Um_f$ is small, although it is expected that the momentum points around which the Dirac Hamiltonians can be derived exist even when $Um_f$ is not small.

Let us analyze Eq. (\ref{H(k)}).
The important point is that all the five alpha matrices which anticommute with each other are used.
To be specific, let us first consider $\mathcal{H}(\tilde{X}^x+\bm{q})$.
We can redefine the alpha matrices because the representation of the matrices is arbitrary.
Redefining such that $\alpha_5\rightarrow\alpha_1$, $\alpha_3\rightarrow -\alpha_3$, and $\alpha_1\rightarrow -\alpha_5$ ($\alpha_5=\alpha_1\alpha_2\alpha_3\alpha_4$) for the alpha matrices, and $t'q_x\rightarrow q_x$, $2\lambda q_y\rightarrow q_y$, and $2\lambda q_z\rightarrow q_z$ for the wave vector \cite{comment1}, we obtain
\begin{align}
\begin{split}
\mathcal{H}(\tilde{X}^x+\bm{q})=q_x\alpha_1+q_y\alpha_2+q_z\alpha_3+\delta t_1'\alpha_4+Um_1\alpha_5.\label{Hx}
\end{split}
\end{align}
In the same manner, $\mathcal{H}(\tilde{X}^y+\bm{q})$ and $\mathcal{H}(\tilde{X}^z+\bm{q})$ can be rewritten as
\begin{align}
\begin{split}
\mathcal{H}(\tilde{X}^y+\bm{q})&=q_x\alpha_1+q_y\alpha_2+q_z\alpha_3+\delta t_1'\alpha_4+Um_2\alpha_5,\\
\mathcal{H}(\tilde{X}^z+\bm{q})&=q_x\alpha_1+q_y\alpha_2+q_z\alpha_3+\delta t_1'\alpha_4+Um_3\alpha_5.\label{Hy-Hz}
\end{split}
\end{align}
We see that all the three effective Hamiltonians above are equivalent.
Hence we can regard the Dirac quasiparticles around the $\tilde{X}^r$ points as the quasiparticles of three flavors characterized by their masses $Um_f$.
Note that the mass of Dirac quasiparticles $\delta t_1$ is renormalized to be $\delta t_1'$ due to the nearest neighbor electron-electron interaction, and that the second mass $Um_f$ is induced by the on-site interaction.

{\it The theta term.---} Here we derive the theta term in the antiferromagnetic insulator phase, in the same way as that of 3D topological insulators is derived.
From the discussion above, we can write down the low-energy effective (Euclidean) action of the system, i.e., the action of the Dirac quasiparticles interacting with an external electromagnetic field $A_\mu$ as
\begin{align}
\begin{split}
S_{\rm AFI}=\int d^4 x\sum_{f=1,2,3}\bar{\psi}_f(x)\left[\gamma_\mu D_\mu-M_f e^{i\kappa_f\gamma_5}\right]\psi_f(x),\label{eff-Action}
\end{split}
\end{align}
where $\psi_f(x)$ is a four-component spinor, $D_\mu=\partial_\mu+ieA_\mu$, $M_f=\sqrt{(\delta t_1')^2+(Um_f)^2}$, $\cos\kappa_f=|\delta t_1'|/M_f$, $\sin\kappa_f=Um_f/M_f$, and we have used the fact that $\alpha_4=\gamma_0$, $\alpha_5=-i\gamma_0\gamma_5$ and $\alpha_j=\gamma_0\gamma_j$ ($j=1,2,3$).
The subscript $f$ denotes the flavor.
Here we have considered the case of $\delta t_1'<0$, namely the system is a normal insulator when the interactions are weak.

We follow the Fujikawa's method \cite{Fujikawa1979,Fujikawa1980} and write down a calculation briefly in what follows.
Let us consider a infinitesimal chiral transformation for each flavor:
\begin{align}
\begin{split}
\psi_f\rightarrow \psi_f'= e^{-i\kappa_f d\phi\gamma_5/2}\psi_f,\ \ \ \ \bar{\psi}_f\rightarrow \bar{\psi}_f'=\bar{\psi}_fe^{-i\kappa_f d\phi\gamma_5/2},
\end{split}
\end{align}
where $\phi\in[0,1]$.
The theta term is generated as a consequence of the chiral anomaly after the transformation.
The partition function is transformed as
\begin{align}
\begin{split}
Z=\int \mathcal{D}[\psi,\bar{\psi}]e^{-S_{\rm AFI}[\psi,\bar{\psi}]}\rightarrow Z'=\int \mathcal{D}[\psi',\bar{\psi}']e^{-S_{\rm AFI}[\psi',\bar{\psi}']}.
\end{split}
\end{align}
The integrands in Eq. (\ref{eff-Action}) is transformed as
\begin{align}
\begin{split}
\bar{\psi}_fM_fe^{i\kappa_f\gamma_5}\psi_f&\rightarrow \bar{\psi}_fM_fe^{i\kappa_f(1-d\phi)\gamma_5}\psi_f,\\
\bar{\psi}_f\gamma_\mu D_\mu\psi_f&\rightarrow \bar{\psi}_f\gamma_\mu D_\mu\psi_f+(i/2)\kappa_f d\phi\partial_\mu (\bar{\psi}_f\gamma_\mu\gamma_5\psi_f).
\end{split}
\end{align}
Then defining the Jacobian $J_f$ which is induced by the chiral transformation for each flavor $\mathcal{D}[\psi_f,\bar{\psi}_f]\rightarrow J_f\mathcal{D}[\psi_f,\bar{\psi}_f]$, the partition function becomes
\begin{align}
\begin{split}
Z'=\int \mathcal{D}[\psi,\bar{\psi}]e^{-S'+\frac{i}{2}\sum_f\kappa_f\int d^4 xd\phi\partial_\mu (\bar{\psi}_f\gamma_\mu\gamma_5\psi_f)+\sum_f\ln J_f},\label{Z-prime}
\end{split}
\end{align}
where
\begin{align}
\begin{split}
S'=\int d^4 x\sum_f\bar{\psi}_f(x)\left[\gamma_\mu D_\mu-M_f e^{i\kappa_f(1-d\phi)\gamma_5}\right]\psi_f(x),
\end{split}
\end{align}
and the Jacobian $J_f$ is calculated to be \cite{Fujikawa1979,Fujikawa1980}
\begin{align}
\begin{split}
J_f=\exp\left[-i\int d^4xd\phi\frac{\kappa_f e^2}{32\pi^2\hbar c}\epsilon^{\mu\nu\rho\lambda}F_{\mu\nu}F_{\rho\lambda}\right].
\end{split}
\end{align}
Here $F_{\mu\nu}=\partial_\mu A_\nu-\partial_\nu A_\mu$ and we have written $\hbar$ and $c$ explicitly.
We repeat this procedure infinite times, i.e., integrate the exponent of Eq. (\ref{Z-prime}) over the variable $\phi$ from $0$ to $1$.
Then we obtain
\begin{align}
\begin{split}
Z'=\int \mathcal{D}[\psi,\bar{\psi}]e^{-S_{\rm NI}+\frac{i}{2}\sum_f\kappa_f\int d^4 x\partial_\mu (\bar{\psi}_f\gamma_\mu\gamma_5\psi_f)-S_\theta},\label{Z-prime2}
\end{split}
\end{align}
where $S_{\rm NI}$ is the action which represents the normal insulator phase in the present case:
\begin{align}
\begin{split}
S_{\rm NI}=\int d^4 x\sum_f\bar{\psi}_f(x)\left[\gamma_\mu D_\mu-M_f\right]\psi_f(x).
\end{split}
\end{align}
This is because the system with negative mass of the Dirac quasiparticles is identified from the $Z_2$ invariant as a normal insulator.
$S_\theta$ is the theta term in the Euclidean spacetime:
\begin{align}
\begin{split}
S_\theta=i\int d^4x\frac{(\sum_f\kappa_f) e^2}{32\pi^2\hbar c}\epsilon^{\mu\nu\rho\lambda}F_{\mu\nu}F_{\rho\lambda}.\label{S_theta}
\end{split}
\end{align}
After dropping the irrelevant surface term [the second term of the exponent in Eq. (\ref{Z-prime2})], we obtain the total action of the system as
\begin{align}
\begin{split}
S_{\rm AFI}=S_{\rm NI}+S_\theta.
\end{split}
\end{align}
Actually $S_\theta$ is also a surface term, since we can rewrite as $\epsilon^{\mu\nu\rho\lambda}F_{\mu\nu}F_{\rho\lambda}=2\epsilon^{\mu\nu\rho\lambda}\partial_\mu(A_\nu F_{\rho\lambda})$.
However, we are now interested in the magnetoelectric response of the system.
Thus we denote the total action as above.
Rewriting the theta term (\ref{S_theta}) in the real time ($t=-i\tau$), we obtain Eq. (\ref{S_theta_realtime}).

The value of $\theta$ in the antiferromagnetic insulator phase is given as $\theta=\sum_f\kappa_f=\sum_f\tan^{-1}(Um_f/|\delta t_1'|)$.
It is known that $\theta=\pi$ (mod $2\pi$) in 3D topological insulators and is $\theta=0$ in normal insulators.
However, $\theta$ can be arbitrary between $0$ and $\pi$ if time-reversal symmetry of the system is broken.
We can obtain the value of $\theta$ in the case of $\delta t_1>0$ in the same manner as above.
Combining both cases, $\theta$ is written as
\begin{align}
\begin{split}
\theta=\frac{\pi}{2}\left[1+{\rm sgn}(\delta t_1)\right]-\sum_{f=1,2,3}\tan^{-1}\left[\frac{Um_f}{\delta t_1(1+V\Delta/t)}\right],\label{Theta}
\end{split}
\end{align}
where the condition that $Um_f\ll 2\lambda$ is required, and we have written $\delta t_1'=\delta t_1(1+V\Delta/t)$ explicitly.
Note that $(1+V\Delta/t)$ is always positive, and thus the value of $\theta$ when $m_f=0$ is determined by the sign of $\delta t_1$.
The region where the value of $\theta$ becomes nonzero is shown in Fig. \ref{Fig2} as the ``axionic SDW''.

Here we compare our analytical result for the value of $\theta$, Eq. (\ref{Theta}), with an exact numerical value calculated by Eq. (\ref{exact-theta1}).
A numerical study on the value of $\theta$ in the Fu-Kane-Mele model on a diamond lattice \cite{Essin2009} indicates the relation $\theta\propto\sum_f\tan^{-1}(Um_f/|\delta t_1'|)$ when $(Um_f/|\delta t_1'|)$ is small.
Thus our result is in qualitative agreement with the exact numerical result.
From our analytical result $\theta=\sum_f\tan^{-1}(Um_f/|\delta t_1'|)$, it is found that $\theta\rightarrow3\pi/2$ in the limit $(Um_f/|\delta t_1'|) \rightarrow \infty$.
The numerical study shows that $\theta$ do not have the $\sum_f\tan^{-1}(Um_f/|\delta t_1'|)$ dependence when $(Um_f/|\delta t_1'|)$ is large, and that $\theta$ takes some value between $0$ and $\pi$  in the limit $(Um_f/|\delta t_1'|) \rightarrow \infty$ \cite{Essin2009}.
This suggests that our analytical result is considered not to be valid when $(Um_f/|\delta t_1'|)$ is large, i.e., $Um_f$ is large.
The difference between our result and the exact numerical result in the region where $Um_f$ is not small can result from that (i) the other contributions to $\theta$ becomes important and unignorable as $Um_f$ becomes larger, and (ii) it might become impossible to derive the Dirac Hamiltonians around the points near the original $X$ points as $Um_f$ becomes larger.

\section{Discussions \label{discussions}}
It should be noted that the theta term is derived only in odd spatial dimensions.
In the Kane-Mele-Hubbard model on the honeycomb lattice at half-filling, which is a two-dimensional analog of the Fu-Kane-Mele-Hubbard model, the antiferromagnetic insulator phase is also realized \cite{Rachel2010,Hohenadler2011,Yamaji2011,Hohenadler2012,Vaezi2012,Wu2012}.
However, the magnetoelectric response which results from the theta term does not appear in that model.

The origin that generates small deviations of the value of $\theta$ from $0$ or $\pi$ in the antiferromagnetic insulator phase of the Fu-Kane-Mele-Hubbard model is the existence of the $\gamma_5$ (or $\alpha_5$ in our notation) term, which breaks time-reversal symmetry, in the low-energy effective action (\ref{eff-Action}).
What we would like to stress here is that we found the appearance of the $\gamma_5$ term in the antiferromagnetic insulator phase.
This is not apparent at first sight of the mean-field Hamiltonian (\ref{SDW-Hamiltonian}).
Expanding the mean-field Hamiltonian around the $\tilde{X}$ points (which are slightly deviated from the original $X$ points) and relabeling the alpha matrices are essential.

Here we mention the relation between the antiferromagnetic insulator phase in our model and the ``Aoki phase'', a phase with broken time-reversal and parity symmetries in lattice QCD.
The Aoki phase is characterized as the phase induced by interactions with the $\gamma_5$ term, i.e., $\langle \bar{\psi}i\gamma_5\psi\rangle\neq 0$ in addition to the usual mass renormalization $\langle \bar{\psi}\psi\rangle$ \cite{Aoki1984}.
It can be seen from the effective Hamiltonian [Eqs. (\ref{Hx}) and (\ref{Hy-Hz})] that the situation in our model is analogous.
Thus it can be said that the antiferromagnetic insulator phase of the extended Fu-Kane-Mele-Hubbard model is a condensed matter analog of the Aoki phase in lattice QCD.
In other words, the Aoki phase in condensed matter in three spatial dimensions can be characterized by the magnetoelectric response which results from the theta term with non-quantized value of $\theta$, i.e., by the axion electrodynamics.
The existence of a similar condensed matter analog of the Aoki phase in three spatial dimensions has been pointed out in a 3D topological insulator with on-site interactions \cite{Sekine2013}.

The Aoki phase has been found in lattice models for QCD such as the Wilson fermions \cite{Aoki1984,Sharpe1998}, the Nambu-Jona-Lasinio model on a lattice \cite{Aoki1994,Azcoiti2013}, and the Gross-Neveu model on a lattice \cite{Aoki1984}.
In the latter two models, the interactions are local.
Namely, from the viewpoint of the form of interactions, it can be said that Hubbard-like models in condensed matter are similar to the Nambu-Jona-Lasinio model and the Gross-Neveu model.
It is known that the existence of the Aoki phase can solve the U(1) problem in QCD.
However, although the importance of the Aoki phase has been confirmed theoretically, the phase is not a realistic phase. 
This is because the phase is an artifact due to the nonzero lattice spacing of lattice QCD \cite{Sharpe1998}, and in addition, the appearance of the phase depends on the value of bare quark mass \cite{Aoki1984}.
On the other hand, in condensed matter, electron systems can be naturally defined on lattices, and the value of the bare mass of Dirac fermions is tunable.
In our model, as mentioned in Sec. \ref{model}, the value is determined by the strength of lattice distortion.
Moreover, experimental searches in condensed matter are possible in principle.
Further investigations of the Aoki phase in condensed matter might enable us to suggest some perception to the field of lattice QCD.
This is an interesting future subject.

Finally let us consider briefly the dynamical behavior of $\theta$ in our model, the dynamical axion field, discussed in Ref. \citen{Li2010}.
When the spins fluctuate, i.e., when $\langle \bm{S}_{A}\rangle=-\langle \bm{S}_{B}\rangle=[m_1+\delta m_1(\bm{r},t)]\bm{e}_x+[m_2+\delta m_2(\bm{r},t)]\bm{e}_y+[m_3+\delta m_3(\bm{r},t)]\bm{e}_z$, we obtain the fluctuation of $\theta$ up to the linear order in $\delta m_f$ as
\begin{align}
\begin{split}
\delta\theta(\bm{r},t)&\approx\sum_{f=1,2,3}\tan^{-1}\left\{U\left[m_f+\delta m_f(\bm{r},t)\right]/|\delta t_1'|\right\}\\
&\quad-\sum_{f=1,2,3}\tan^{-1}\left(Um_f/|\delta t_1'|\right)\\
&\approx U/|\delta t_1'|\sum_{f}\delta m_f(\bm{r},t).
\end{split}\label{dynamical-axion}
\end{align}
This equation suggests that the dynamical axion field can be realized by the fluctuations of the spins, i.e., the spin-wave excitations, as in the case of Ref. \citen{Li2010}.
An advantage of our analytical derivation of the expression of $\theta$ is that we can see the realization of the dynamical axion field immediately, as Eq. (\ref{dynamical-axion}).
In the case where we use expressions (\ref{exact-theta1}) or (\ref{exact-theta2}), it will not easy to notice the realization in our model.

\section{Summary}
In summary, we have studied the ground state and the topological magnetoelectric response described by the theta term, in the extended Fu-Kane-Mele-Hubbard model on a diamond lattice at half-filling, within the mean-field approximation.
The mean-field phase diagram was presented.
It was found that the transition from the normal insulator (or topological insulator) phase to the antiferromagnetic insulator phase is of the second-order.
We obtained the Dirac-like low-energy effective Hamiltonian in the antiferromagnetic insulator phase.
We found that there exists the $\gamma_5$ term in the effective Hamiltonian.
This antiferromagnetic insulator phase is different from conventional one, and can be regarded as a condensed matter analog of a symmetry broken phase in lattice QCD.
We derived the theta term by following the Fujikawa's method and obtained the analytical value of $\theta$.
We have proposed a concrete model to describe the axion electrodynamics in the antiferromagnetic insulator phase.
The dynamical axion field can be induced by the fluctuation of the order parameter.
In our model, the interplay of spin-orbit coupling and electron correlation results in the emergence of the topological magnetoelectric response and the realization of the dynamical axion field.

\section*{Acknowledgments}
A.S. is supported by the JSPS Research Fellowship for Young Scientists.
This work was supported by Grant-in-Aid for Scientific Research (No. 25103703, No. 26107505 and No. 26400308) from the Ministry of Education, Culture, Sports, Science and Technology (MEXT), Japan.



\begin{thebibliography}{99}
\bibitem{comment1}
When interacting with an electromagnetic field, the vector potential $\bm{A}$ is rescaled as well as the spatial coordinate $\bm{x}$.
Then we see that the theta term (\ref{S_theta}) is invariant under rescaling.
Namely, the value of $\theta$ does not depend on the values of $\lambda$ and $t$ in the low-energy limit.
\end{thebibliography}
\end{document}